\documentclass[conference]{IEEEtran}
\IEEEoverridecommandlockouts
\usepackage{cite,balance}
\usepackage{amsmath,amssymb,amsfonts}
\usepackage{algorithmic}
\usepackage[ruled,vlined]{algorithm2e}
\usepackage{graphicx}
\usepackage{textcomp}
\usepackage{xcolor}
\usepackage{optidef}
\usepackage{tikz}
\usepackage{pgfplots}
\usepackage{verbatim}
\usepackage{mathrsfs}

\def\BibTeX{{\rm B\kern-.05em{\sc i\kern-.025em b}\kern-.08em
    T\kern-.1667em\lower.7ex\hbox{E}\kern-.125emX}}
\begin{document}

\title{Joint Optimization of Execution Latency and Energy Consumption for Mobile Edge Computing with Data Compression and Task Allocation \\
}

\author{\IEEEauthorblockN{Minh Hoang Ly\textsuperscript{1}, Thinh Quang Dinh\textsuperscript{2}, Ha Hoang Kha\textsuperscript{1}}
\IEEEauthorblockA{
\textsuperscript{1}\textit{Ho Chi Minh City University of Technology, VNU-HCM, Vietnam} \\
hoang.ly-minh@student-cs.fr, hhkha@hcmut.edu.vn \\
\textsuperscript{2}\textit{Trusting Social, Vietnam} \\
thinh.dinh@trustingsocial.com
}


}

\maketitle

\begin{abstract}
This paper studies the mobile edge offloading scenario consisting of one mobile device (MD) with multiple independent tasks and various remote edge devices. In order to save energy, the user's device can offload the tasks to available access points for edge computing. Data compression is exploited to reduce offloaded data size prior to wireless transmission to minimize the execution latency. The problem of jointly optimizing the task allocation decision and the data compression ratio to minimize the total tasks' execution latency and the MD's energy consumption concurrently is proposed. We show that the design problem is a non-convex optimization one but it can be recasted into a convex one through a semidefinite relaxation (SDR) based approach. Numerical simulations demonstrate the outperformance of the proposed scheme compared to the benchmark one.
\end{abstract}



\section{Introduction}
Mobile devices have been playing a vital role in today’s world. Various applications, ranging from voice communication, web surfing to navigation are available on a single mobile phone \cite{b1}. At the same time, there is an increasing number of computation-intensive applications providing smart services which mobile devices (MDs) can hardly support due to limited computing ability and battery capacity \cite{b2}. Hence, an energy efficiency-based data processing approach is critical for battery-powered devices. Mobile cloud computing increases the capabilities of MDs to enhance user experience and, thus, can be viewed as a potential solution \cite{b3,b4}. Since there are much higher computation and storage resources on the cloud servers, migrating demanding tasks to clouds can eliminate the burden of computing, storage, and computation energy on the MDs. A variety of offloading frameworks allow the MDs to partition their applications into executable tasks which can be processed on cloud servers with various hardware architectures \cite{b4}.

Although adequate cloud resources can be utilized to support MDs to collect, store and process data, the interaction between the cloud and MDs is challenging. For example, cloud servers are commonly geographically far from MDs, which results in high communications latency. The trade-off between energy savings and computing efficiency also affects user experience. In addition, communication delay and energy cost between MDs and the cloud are not negligible. Moreover, finite backhaul capacity is a bottleneck of remote cloud computing \cite{b2},\cite{b3}. Mobile edge computing (MEC) could be the key to solve the problem. Unlike remote cloud computing, MDs can offload computing tasks to a nearby MEC server to reduce huge backhaul latency caused by offloading to a remote cloud. Consequently, if small cell access points (AP) are capable of processing tasks for MDs, this paradigm can make a reduction in the application response time \cite{b2},\cite{b4}. In 2009, the authors in \cite{b5} introduced the idea of bringing both computational and communication capacities close to users. However, in all works to date, an MD can only associate with either a single remote server or an edge device. With the dense deployment of APs in upcoming networks \cite{b6}, we can use this diversity by considering a scenario when an MD can offload to multiple nearby APs with computational capacities (Computing Access Point or CAP), instead of only a single CAP.

On the other hand, data compression applied in modern applications to reduce the data size of text, audio, image, and other types is ubiquitous \cite{b7},\cite{b8}. By this way, data compressing can save storage space or lower data transmission time in wireless networks \cite{b2}. Moreover, implementing data compression in MEC systems would help save transmission energy since energy consumption for transmitting data is normally significantly higher than that of computing \cite{b2}. For the reason that executing data compression algorithms takes time and energy, there is a cost of applying data compression on all data in MEC systems. Thus, an amount of offloaded task data to be compressed together with task allocation and offloading decision should be optimized. In this context, extensive works have focused on investigating energy or latency minimizing problems but few works jointly considered data compression and MEC \cite{b2},\cite{b9}. Additionally, computational offloading affects both the tasks’ execution latency and energy consumption of the MD. However, most of the earlier work either concentrated only on execution latency \cite{b10},\cite{b11} or energy consumption \cite{b12},\cite{b13}. To the best of our knowledge, no previous work has studied in applying data compression in MEC systems to jointly minimize execution latency and energy consumption.

In this work, we investigate the case when a single MD is able to allocate tasks to multiple CAPs. We introduce a framework where the MD minimizes a weighted sum of the cost of its tasks’ execution latency and energy consumption by jointly optimizing the task allocation decision and the task compression ratio. By exploiting this flexibility in terms of allocation decision and compression ratio, we can improve the MD’s task latency performance and energy consumption. We show that the associated problem can be formulated as a non-convex quadratically constrained quadratic program (QCQP), which is, in general, NP-hard. To tackle this problem, an efficient semidefinite relaxation (SDR) algorithm \cite{b14} and a randomization mapping method is proposed. Simulation results validate the advantages of our approach.

\section{System model and problem formulation}
\begin{figure}[htbp]
	\centerline{\includegraphics[width=0.45\textwidth]{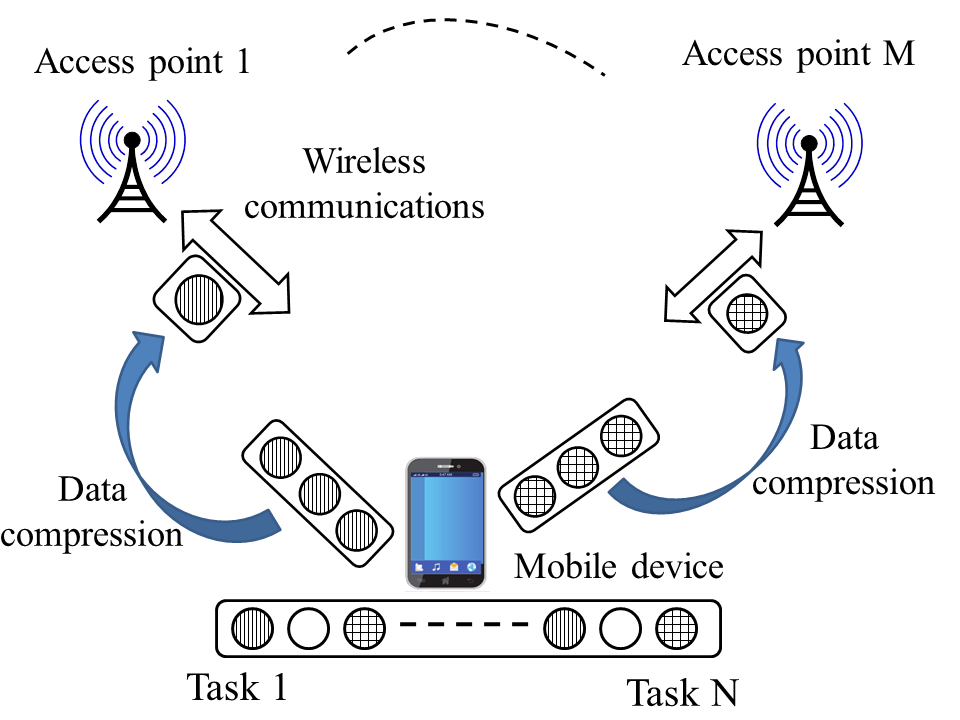}}
	\caption{Offloading framework with a single MD and multiple CAPs.}
	\label{fig: Hinh1}
\end{figure}

Consider the framework that consists of an MD with $N$ independent tasks. Each of these tasks can be locally processed by the MD's CPU or offloaded to a CPU of  $M$ wireless CAPs as shown in Fig. \ref{fig: Hinh1}. Before offloading, the potentially offloaded tasks can be compressed to reduce the transmission energy. The set of tasks and the set of CPUs are denoted by $\mathcal{N} = \{1,..., N\}$ and $\mathcal{M} = \{0,..., M\}$, respectively, where CPU $0$ is the MD's CPU {\footnote {Later on, unless otherwise stated, we include MD's CPU whenever we mention CPUs}}. The MD has to make the decision about whether tasks would be remotely or locally processed.

Every task $i, i \in \mathcal{N}$, is represented by a tuple $\{\alpha_i, \beta_i, \omega_i\}$ where $\alpha_i$ is the input data size (in bits), $\beta_i$ is the output data size (in bits), and $\omega_i$ is the number of CPU cycles that are required to process the task, respectively. We assume that the amount of data to be processed, which is of irregular sizes, is known beforehand \cite{b15}. Each of the CAPs offers the MD with fixed service rate (CPU frequency) $r_k$ (cycles/sec or Hertz). The downlink and uplink rates between the CAPs and user are denoted as $C_k^{\text{DL}}$ and $C_k^{\text{UL}}$ (bits/sec), respectively. In this paper, it is assumed that CAPs have full channel state information (CSI) and, thus, $C_k^{\text{UL}}$ and $C_k^{\text{DL}}$ are known to the MD. Therefore, the effective rate $C_k^{\text{DL}}$ and $C_k^{\text{UL}}$ during the whole period can be approximated by their mean values. An indicator $x_{ik}, \forall i \in \mathcal{N}, k \in \mathcal{M}$ is utilized to denote the task allocation,
$$
x_{ik} = \begin{cases} 1,  \text{   if task $i$ is assigned to CPU $k$,}  \\ 0,  \text{   otherwise.} \end{cases}
$$
Define $\textbf{X} = \{x_{ik}\} \in \{0, 1\}^{N\times(M+1)}$ the $\textit{task}$ $\textit{allocation}$ $\textit{matrix}$, and $\textbf{x} = [\textbf{x}_0^T, \textbf{x}_1^T,...,\textbf{x}_k^T,...\textbf{x}_M^T]^T$, where $\textbf{x}_k = [x_{1k}, x_{2k},..., x_{Nk}]^T$, i.e., the $k^{th}$ column vector of $\textbf{X}$.

We assume that when a task is about to be uploaded to a CAP, all the fraction of the task (the whole task) is offloadable, \cite{b4} and prior to transmission, the task can be compressed down to an arbitrary size. It is also assumed that lossless data compression methods are manipulated so that original data can be perfectly reconstructed at the MEC server \cite{b2}. Let $J^{\text{C}}$ (cycles/bit) and $E^{\text{C}}$ (Joules/cycle) be the computation for compressing 1-bit of task data and energy consumption per CPU cycle at the MD, respectively. We assume that $\gamma \in [0, 1]$ fraction of $\textit{all}$ $\textit{offloaded}$ $\textit{task}$ is compressed before offloading, i.e. the size of compressed data (after the compression step) is equivalent to $(1-\gamma)$ times the size of original data. The time delay for task $i$'s local compression is $t_i^{\text{Compr}} = \frac{\alpha_iJ^{\text{C}}\gamma}{r_0}$ \cite{b2}. The time delay for task $i$'s decompression at $k^{th}$ CAP is $t_{ik}^{\text{Decompr}} = \frac{\alpha_iJ^{\text{C}}\gamma}{r_k}$ \cite{b7}.

We define $g_{ik}^{\text{DL}}$, $g_{ik}^{\text{UL}}$ and $g_{ik}^{\text{Comp}}$ as the downlink, uplink and computation latency, respectively, when offload task $i$ to CPU $k$, where $g_{ik}^{\text{UL}} = \frac{\alpha_i}{C_k^{\text{UL}}}$, $g_{ik}^{\text{DL}} = \frac{\beta_i}{C_k^{\text{DL}}}$ and $g_{ik}^{\text{Comp}} = \frac{\omega_i}{r_k}$. Therefore, the total compression latency $T_{k}^{\text{Compr}}$, upload latency $T_{k}^{\text{UL}}$, decompression latency $T_{k}^{\text{Decompr}}$, computation latency $T_{k}^{\text{Comp}}$ and download latency $T_{k}^{\text{DL}}$ are given by $T_{k}^{\text{Compr}} = \dfrac{\sum_{i \in \mathcal{N}}x_{ik}\alpha_iJ^{\text{C}}\gamma}{r_0}$, $T_{k}^{\text{UL}} = \dfrac{\sum_{i \in \mathcal{N}}x_{ik}\alpha_i(1-\gamma)}{C_k^{\text{UL}}}$, $T_{k}^{\text{Decompr}} = \dfrac{\sum_{i \in \mathcal{N}}x_{ik}\alpha_iJ^{\text{C}}\gamma}{r_k}$, $T_{k}^{\text{Comp}} = \dfrac{\sum_{i \in \mathcal{N}}x_{ik}\omega_i}{r_k}$ and $T_{k}^{\text{DL}} = \dfrac{\sum_{i \in \mathcal{N}}x_{ik}\beta_i}{C_k^{\text{DL}}}$, respectively. In this paper, for simplicity, we assume independent and non-overlapped steps at CAPs, meaning that a CAP only starts to process tasks after receiving all. The execution latency of the batch uploaded to CPU $k$, $\forall k \in \mathcal{M}$ is computed as
$T_k = T_{k}^{\text{Compr}}+ T_{k}^{\text{UL}} + T_{k}^{\text{Decompr}}+ T_{k}^{\text{Comp}} + T_{k}^{\text{DL}}$$ $$=\sum_{i \in \mathcal{N}}x_{ik}\bigg(\dfrac{\alpha_iJ^{\text{C}}\gamma}{r_0} + \dfrac{\alpha_i(1-\gamma)}{C_k^{\text{UL}}} + \dfrac{\alpha_iJ^{\text{C}}\gamma}{r_k} + \dfrac{\omega_i}{r_k} + \dfrac{\beta_i}{C_k^{\text{DL}}} \bigg)$
$=\sum_{i \in \mathcal{N}}x_{ik}\text{G}_{ik}$, where $\text{G}_{ik} = \dfrac{\alpha_iJ^{\text{C}}\gamma}{r_0} + \dfrac{\alpha_i(1-\gamma)}{C_k^{\text{UL}}} + \dfrac{\alpha_iJ^{\text{C}}\gamma}{r_k} + \dfrac{\omega_i}{r_k} + \dfrac{\beta_i}{C_k^{\text{DL}}}$. For the local processing case, there are no uplink, downlink as well as compression and decompression latency, i.e., $\text{G}_{i0} = \frac{\omega_i}{r_0}$.

The energy consumption of the MD is mostly contributed by computational processing, local compression and (wireless) transmission.
\begin{itemize}
\item The computational energy consumption of the MD is
$$E^{\text{Comp}} = P^{\text{Comp}}\sum_{i \in \mathcal{N}}x_{i0}\text{G}_{i0},$$
where $P^{\text{Comp}}$ is the computational power of the MD.
\item Local compression energy consumption is defined as \cite{b2}
$\begin{array} {lcl} E^{\text{Compr}} & = & J^{\text{C}}E^{\text{C}}\sum_{k \in \mathcal{M}\texttt{\symbol{92}}\{0\}}\sum_{i \in \mathcal{N}}x_{ik}\alpha_i\gamma \\ & = & P^{\text{Compr}}\sum_{k \in \mathcal{M}\texttt{\symbol{92}}\{0\}}\sum_{i \in \mathcal{N}}x_{ik}\alpha_i\gamma \end{array}$,\\
where $P^{\text{Compr}} = J^{\text{C}}E^{\text{C}}$ (Joules/bit) is the power consumption for 1-bit compression at the MD.
\item Transmission energy consumption is given as
$E^{\text{TR}} = P^{\text{Tx}}\sum_{k \in \mathcal{M}\texttt{\symbol{92}}\{0\}}\sum_{i \in \mathcal{N}}x_{ik}\dfrac{\alpha_i(1-\gamma)}{C_k^{\text{UL}}} + P^{\text{Rx}}\sum_{k \in \mathcal{M}\texttt{\symbol{92}}\{0\}}\sum_{i \in \mathcal{N}}x_{ik}\dfrac{\beta_i}{C_k^{\text{DL}}}$ $= \sum_{k \in \mathcal{M}\texttt{\symbol{92}}\{0\}}\sum_{i \in \mathcal{N}}\Big(P^{\text{Tx}}x_{ik}g_{ik}^{\text{UL}}(1-\gamma) + P^{\text{Rx}}x_{ik}g_{ik}^{\text{DL}}\Big)$, where $P^{\text{Tx}}$ and $P^{\text{Rx}}$, which, regarded as constants, are the transmitting and receiving power, respectively.
\end{itemize}
The metrics of the main objective functions can be represented as follows:
\begin{itemize}
	\item Execution latency: Since CAPs are able to process tasks in parallel \cite{b4}, the execution latency is defined as
	$$t(\mathcal{\textbf{X}}, \gamma) = \text{max}_{k \in \mathcal{M}}T_k.$$
	\item Total energy consumption: The total energy consumption of the MD is given by
	$$e(\mathcal{\textbf{X}}, \gamma) = E^{\text{Comp}} + E^{\text{Compr}} + E^{\text{TR}}.$$
\end{itemize}
In the subsequent sections, we jointly minimize the weighted sum of execution latency and energy consumption by investigating the trade-off between the two objectives and defining the joint objective function as
\begin{equation}
\psi(\textbf{X}, \gamma) = \lambda_tt(\textbf{X}, \gamma) + \lambda_ee(\textbf{X}, \gamma),
\end{equation}
where $\lambda_t, \lambda_e \in [0, 1]$ are scalar weights. These weights itself demonstrate the correlation and relative significance between the latency and energy. The problem can be formulated as Problem $\mathscr{P}1$:
\begin{subequations}
	\begin{align}
	\nonumber		\mathscr{P}1:			\min_{\textbf{X}, \gamma} ~& \psi(\textbf{X}, \gamma)\\
	\text{s.t} ~~& {\sum_{k \in \mathcal{M}}x_{ik} = 1}, \quad \forall  i \in \mathcal {N}, \\
	& x_{ik}\in \{0, 1\},\\
	& \gamma\in [0, 1].
	\end{align}
\end{subequations}	

$\mathscr{P}1$ is an NP-hard problem \cite{b4}.
\section{Task offloading optimization solution}
To find the optimal solution of $\mathscr{P}1$, we introduce a new variable $t$ with additional constraint $t = \text{max}_{k \in \mathcal{M}}\big(\sum_{i \in \mathcal{N}}x_{ik}\text{G}_{ik}\big)$ to transform $\mathscr{P}1$ to a new problem $\mathscr{P}2$
\begin{subequations}
	\begin{align}
	\nonumber		\mathscr{P}2:			\min_{\textbf{X}, \gamma} ~& \psi(\textbf{X}, \gamma)=\lambda_tt + \lambda_e\big(E^{\text{Comp}} + E^{\text{Compr}} + E^{\text{TR}} \big)\\
	\text{s.t} ~~& \sum_{i \in \mathcal{N}}x_{ik}\text{G}_{ik}\leq t, \quad \forall  k \in \mathcal {M}, \\
	& \sum_{k \in \mathcal{M}}x_{ik} = 1, \forall i \in \mathcal{N} \\
	& x_{ik}\in \{0, 1\},\\
	& \gamma\in [0, 1].
	\end{align}
\end{subequations}	

\subsection{Quadratic Program Transformation}
To convert the optimization problem $\mathscr{P}2$ into a semidefinite program (SDP), we transform it to a quadratically constrained quadratic program (QCQP). Define $\textbf{y} = [\textbf{x}^T, \gamma, t]^T$ and $Q = NM + N$, the objective function can be rewritten as
$\psi(\textbf{X}, \gamma) = \lambda_tt + \lambda_eP^{\text{Comp}}\sum_{i \in \mathcal{N}}x_{i0}\text{G}_{i0} + \lambda_eP^{\text{Compr}}\sum_{k \in \mathcal{M}\texttt{\symbol{92}}\{0\}}\sum_{i \in \mathcal{N}}x_{ik}\alpha_i\gamma + \\
\lambda_e\sum_{k \in \mathcal{M}\texttt{\symbol{92}}\{0\}}\sum_{i \in \mathcal{N}}\Big(P^{\text{Tx}}x_{ik}g_{ik}^{\text{UL}}(1-\gamma) + P^{\text{Rx}}x_{ik}g_{ik}^{\text{DL}}\Big) = \textbf{y}^T(\lambda_eP^{\text{Compr}}\textbf{A}_1 - P^{\text{Tx}}\lambda_e\textbf{A}_2)\textbf{y} + \textbf{b}_0^T\textbf{y} = \psi(\textbf{y})$,
where
$$\textbf{A}_1 = \begin{bmatrix}
\textbf{0}_{Q \times Q}  & & \textbf{0}_{N\times 1} & \textbf{0}_{(Q+1)\times 1} \\
& & \dfrac{\boldsymbol{\alpha}_M}{2}  &  \\
\textbf{0}_{1\times N} & \dfrac{{\boldsymbol{\alpha}_M}^T}{2} & 0 & \\
\textbf{0}_{1\times(Q+1)} & & & 0 & \\
\end{bmatrix}, $$
$\boldsymbol{\alpha} = [\alpha_1, \alpha_2,..., \alpha_N]^T$, $\boldsymbol{\alpha}_M = [\underbrace{\boldsymbol{\alpha}^T, \boldsymbol{\alpha}^T,...,\boldsymbol{\alpha}^T}_{\text{M times}}]^T$,
$$\textbf{A}_2 = \begin{bmatrix}
\textbf{0}_{Q\times Q}  & & \textbf{0}_{N\times 1} & \textbf{0}_{(Q+1)\times 1} \\
& & \dfrac{\boldsymbol{g}^{\text{UL}}}{2}  &  \\
\textbf{0}_{1\times N} & \dfrac{{\boldsymbol{g}^{\text{UL}}}^T}{2} & 0 & \\
\textbf{0}_{1\times(Q+1)} & & & 0 & \\
\end{bmatrix},$$
$\boldsymbol{g}^{\text{UL}} = \Big[{\boldsymbol{g}_1^{\text{UL}}}^T, {\boldsymbol{g}_2^{\text{UL}}}^T,..., {\boldsymbol{g}_M^{\text{UL}}}^T\Big]^T$ and $\boldsymbol{g}^{\text{DL}} = \Big[{\boldsymbol{g}_1^{\text{DL}}}^T, {\boldsymbol{g}_2^{\text{DL}}}^T,..., {\boldsymbol{g}_M^{\text{DL}}}^T\Big]^T$ where
$\boldsymbol{g}_k^{\text{UL}} = \Big[g_{1k}^{\text{UL}}, g_{2k}^{\text{UL}},..., g_{Nk}^{\text{UL}}\Big]^T$, $\boldsymbol{g}_k^{\text{DL}} = \Big[g_{1k}^{\text{DL}}, g_{2k}^{\text{DL}},..., g_{Nk}^{\text{DL}}\Big]^T, k \in \mathcal{M}\texttt{\symbol{92}}\{0\}$,

$\textbf{b}_0 = \Big[\lambda_eP^{\text{Comp}}\boldsymbol{\ell}_0^T, \lambda_e\textbf{b}_0'^T, 0, \lambda_t\Big]^T$, $\textbf{b}_0' = P^{\text{Tx}}\boldsymbol{g}^{\text{UL}} + P^{\text{Rx}}\boldsymbol{g}^{\text{DL}}$, $\boldsymbol{\ell}_0 = \Big[\dfrac{\omega_1}{r_0}, \dfrac{\omega_2}{r_0},...,\dfrac{\omega_N}{r_0}\Big]^T$.

We now examine the execution latency constraint
$$T_k = \sum_{i \in \mathcal{N}}x_{ik}\text{G}_{ik} \leq t, \forall k \in \mathcal{M}.$$
Note that $G_{ik} = G_{ik}(\gamma)$ is an affine function of $\gamma \in [0, 1]$. Thus, we have $G_{ik}(\gamma) \leq \text{max}\{G_{ik|\gamma = 0}; G_{ik|\gamma = 1}\}$ or equivalently $t \geq \text{max}\{T_{k|\gamma = 0}; T_{k|\gamma = 1}\}$. These constraints, together with constraint (3a) can be rewritten as:

\begin{equation}
\sum_{i \in \mathcal{N}}x_{i0}\dfrac{\omega_i}{r_0} \leq t,
\end{equation}
\begin{equation}
\sum_{i \in \mathcal{N}}x_{ik}\bigg(\dfrac{\alpha_i}{C_k^{\text{UL}}} +  \dfrac{\omega_i}{r_k} + \dfrac{\beta_i}{C_k^{\text{DL}}} \bigg) \leq t, \forall k \in \mathcal{M}\texttt{\symbol{92}}\{0\},
\end{equation}
\begin{equation}
\sum_{i \in \mathcal{N}}x_{ik}\bigg(\alpha_iJ^{\text{C}}\Big(\dfrac{1}{r_0} + \dfrac{1}{r_k}\Big) + \dfrac{\omega_i}{r_k} + \dfrac{\beta_i}{C_k^{\text{DL}}} \bigg) \leq t, \forall k \in \mathcal{M}\texttt{\symbol{92}}\{0\}.
\end{equation}
Let $\textbf{D} = \{\text{D}_{ik}\} = \dfrac{\alpha_i}{C_k^{\text{UL}}} +  \dfrac{\omega_i}{r_k} + \dfrac{\beta_i}{C_k^{\text{DL}}}$ where $i \in \mathcal{N}, k \in \mathcal{M}\texttt{\symbol{92}}\{0\}$; $\textbf{D}_k = [\text{D}_{1k}, \text{D}_{2k},..., \text{D}_{Nk}]^T$, i.e., the $k^{th}$ column of $\textbf{D}$ and $\textbf{G}_0 = [\text{G}_{10}, \text{G}_{20},..., \text{G}_{N0}]^T$. Then constraints (4) and (5) are equivalent to $$\textbf{A}_3\textbf{y} \leq \textbf{0}_{(M+1)\times1},
$$
where $\textbf{A}_3 = \begin{bmatrix}
\textbf{G}_{0}^T  & \textbf{0}_{1\times N} & \textbf{0}_{1\times N} & \cdots & \textbf{0}_{1\times N} & 0 & -1\\
\textbf{0}_{1\times N} & \textbf{D}_{1}^T & \textbf{0}_{1\times N} & \cdots & \textbf{0}_{1\times N} & 0 & -1 \\
\textbf{0}_{1\times N} & \textbf{0}_{1\times N} & \textbf{D}_{2}^T  & \cdots & \textbf{0}_{1\times N} & 0 & -1 \\
\vdots & \vdots & \vdots & \ddots & \vdots & \vdots & \vdots \\
\textbf{0}_{1\times N} & \textbf{0}_{1\times N} & \textbf{0}_{1\times N} & \cdots & \textbf{D}_{M}^T & 0 & -1 \\
\end{bmatrix}.$

Similarly, denote $\textbf{E} = \{\text{E}_{ik}\} = \alpha_iJ^{\text{C}}\Big(\dfrac{1}{r_0}+\dfrac{1}{r_k}\Big) +  \dfrac{\omega_i}{r_k} + \dfrac{\beta_i}{C_k^{\text{DL}}},$ where $i \in \mathcal{N}, k \in \mathcal{M}\texttt{\symbol{92}}\{0\}$ and $\textbf{E}_k = [\text{E}_{1k}, \text{E}_{2k},..., \text{E}_{Nk}]^T$, i.e., the $k^{th}$ column of $\textbf{E}$, then (6) can be rewritten as
$$\textbf{A}_4\textbf{y} \leq \textbf{0}_{M\times 1},
$$
where $\textbf{A}_4 = \begin{bmatrix}
\textbf{0}_{1\times N} & \textbf{E}_{1}^T  & \textbf{0}_{1\times N} & \cdots & \textbf{0}_{1\times N} & 0 & -1 \\
\textbf{0}_{1\times N} & \textbf{0}_{1\times N} & \textbf{E}_{2}^T  & \cdots & \textbf{0}_{1\times N} & 0 & -1 \\
\vdots & \vdots & \vdots & \ddots & \vdots & \vdots & \vdots \\
\textbf{0}_{1\times N} & \textbf{0}_{1\times N} & \textbf{0}_{1\times N} & \cdots & \textbf{E}_{M}^T & 0 & -1 \\
\end{bmatrix}.$

Finally, the binary constraint (3c) can be replaced by a quadratic one:
$$x_{ik} \in \{0, 1\} \leftrightarrow x_{ik}(x_{ik}-1) = 0,$$
and constraint (3d) can be replaced by
$$\gamma \in [0, 1] \leftrightarrow \gamma(\gamma-1) \leq 0.$$

Therefore, problem $\mathscr{P}$2 is equivalent to $\mathscr{P}$3:
\begin{subequations}
	\begin{align}
	\nonumber		\mathscr{P}3:			\min_{\textbf{y}} ~& \psi(\textbf{y})=\textbf{y}^T(\lambda_eP^{\text{Compr}}\textbf{A}_1 - P^{\text{Tx}}\lambda_e\textbf{A}_2)\textbf{y} + \textbf{b}_0^T\textbf{y} \\
	\text{s.t} ~~& \textbf{A}_3\textbf{y} \leq \textbf{0}_{(M+1)\times 1}, \\
	& \textbf{A}_4\textbf{y}\leq \textbf{0}_{M\times 1}, \\
	& \textbf{A}_5\textbf{y}= \textbf{1}_{N\times 1},\\
	& \textbf{y}^T\text{diag}(\textbf{u}_p)\textbf{y} - \textbf{u}_p^T\textbf{y} = 0, \quad    p = 1, 2,..., Q, \\
	& \textbf{y}^T\text{diag}(\textbf{u}_{Q+1})\textbf{y} - \textbf{u}_{Q+1}^T\textbf{y} \leq 0.
	\end{align}
\end{subequations}	

where $\textbf{A}_5 = [\textbf{I}_{N, 0}, \textbf{I}_{N, 1},...,\textbf{I}_{N, M}, \textbf{0}_{N\times1}, \textbf{0}_{N\times1}]$, $\textbf{I}_{N, k}$ is a $N$-dimensional identity matrix, $\forall k \in \mathcal{M}$ and $\textbf{u}_p$ is a $(Q+2)\times1$ unit vector with the $p^{\text{th}}$ entry being 1. Here, constraint (3a) is equivalent to constraints (7a) and (7b) while constraints (3b), (3c), (3d) are equivalent to constraints (7c), (7d) and (7e), respectively.

\subsection{Semidefinite Relaxation Approach}
In this section, we develop an algorithm to tackle problem $\mathscr{P}$3. According to \cite{b16}, by adding the rank one constraint, a QCQP problem can be recast a semidefinite program, which can be efficiently solved in polynomial time using a standard SDP solver \cite{b3},\cite{b17}.

Define $\textbf{Z} = \begin{bmatrix}
	\textbf{y} \\
	1 \\
\end{bmatrix}[\textbf{y}^T  \  1]$ and $Q_2 = Q+2 = MN + N +2$,
$\mathscr{P}$3 is homogenized to $\mathscr{P}$4
\begin{subequations}
	\begin{align}
	\nonumber		\mathscr{P}4:			\min_{\textbf{Z}} ~& \text{Tr}(\textbf{B}_0\textbf{Z}) \\
	\text{s.t} ~~& \text{Tr}(\textbf{H}_h\textbf{Z}) \leq 0, \quad    h = 1, 2,..., M+1, \\
	& \text{Tr}(\textbf{J}_j\textbf{Z}) \leq 0, \quad    j = 1, 2,..., M, \\
	& \text{Tr}(\textbf{G}_p\textbf{Z}) = 1, \quad    p = 1, 2,..., N,\\
	& \text{Tr}(\textbf{K}_q\textbf{Z}) = 0, \quad    q = 1, 2,..., Q, \\
	& \text{Tr}(\textbf{K}_q\textbf{Z}) \leq 0, \quad    q = Q+1, \\
	& \textbf{Z} \succeq 0, \text{rank}(\textbf{Z}) = 1, z_{Q+3, Q+3} = 1.
	\end{align}
\end{subequations}	

where \newline

$\textbf{B}_0 = \begin{bmatrix}
\textbf{A}_6 & \dfrac{1}{2}\textbf{b}_0 \\
\dfrac{1}{2}\textbf{b}_0^T & 0
\end{bmatrix}$,
$\quad \textbf{A}_6 = \lambda_eP^{\text{Compr}}\textbf{A}_1 - \lambda_eP^{\text{Tx}}\textbf{A}_2$,
$\textbf{H}_h = \begin{bmatrix}
\textbf{0}_{Q_2\times Q_2} & \dfrac{\textbf{A}_{3, h}^T}{2} \\
\dfrac{\textbf{A}_{3, h}}{2} & 0
\end{bmatrix}$,
$\quad \textbf{J}_j = \begin{bmatrix}
\textbf{0}_{Q_2\times Q_2} & \dfrac{\textbf{A}_{4, j}^T}{2} \\
\dfrac{\textbf{A}_{4, j}}{2} & 0
\end{bmatrix}$,
\newline
$\textbf{G}_p = \begin{bmatrix}
\textbf{0}_{Q_2\times Q_2} & \dfrac{\textbf{A}_{5, p}^T}{2} \\
\dfrac{\textbf{A}_{5, p}}{2} & 0
\end{bmatrix}$,
$\quad \textbf{K}_q = \begin{bmatrix}
\text{diag}(\textbf{u}_q) & \dfrac{-\textbf{u}_q}{2} \\
\dfrac{-\textbf{u}_q^T}{2} & 0
\end{bmatrix}.$

Moreover, $z_{Q+3, Q+3}$ is the element at row $Q+3$ and column $Q+3$ of the matrix $\textbf{Z}$; $\textbf{A}_{3, h}, \textbf{A}_{4, j}$ and $\textbf{A}_{5, p}$ are the $h^{\text{th}}$, $j^{\text{th}}$ and $p^{\text{th}}$ row of the matrix $\textbf{A}_3, \textbf{A}_4$ and $\textbf{A}_5$, respectively. Constraints (7a)-(7e) correspond to constraints (8a)-(8e), respectively.

Constraint $\text{rank}(\textbf{Z}) = 1$ is the only non-convex constraint in problem $\mathscr{P}$4 while other constraints and the objective function are convex. Let $\textbf{Z}^*$ is the optimal solution of problem $\mathscr{P}$4 without the rank constraint found by the convex programming solver. If $\textbf{Z}^*$ is of rank 1, then the found solution is optimal. If $\text{rank}(\textbf{Z}^*) \neq 1$, we propose an algorithm to obtain an near-optimal solution for problem $\mathscr{P}$1 based on Gaussian randomization. The main ideas of the algorithm are as follows:
\begin{itemize}
	\item Generate randomly $L$ feasible solutions of $\mathscr{P}$1 based on a multivariate Gaussian distribution having zero mean and $\textbf{Z}'$ as the covariance matrix, where $\textbf{Z}'$ is the $Q\times Q$ upper-left sub-matrix of $\textbf{Z}^*$;
	\item Determine the solution that minimizes the objective function in $\mathscr{P}$1.
\end{itemize}

Once the sub-matrix $\textbf{Z}'$ is extracted from $\textbf{Z}$, we can generate $L$ random $(MN+N)\times 1$ vectors where the $l^\text{th}$ vector denoted by $\hat{\textbf{\text{x}}}^{(l)}$ is based on $\hat{\textbf{\text{x}}}^{(l)} \sim \mathcal{N}(0_{(MN+N)\times 1}, \textbf{Z}'),  l = 1,..., L$. Note that the number of elements in $\hat{\textbf{\text{x}}}^{(l)}$ is equal to the number of elements in $\textbf{\text{x}}$. Here, we need to recover the binary characteristic of $\hat{\textbf{\text{x}}}^{(l)}$ to make it a valid solution of $\mathscr{P}$1. To satisfy constraint (2a), we recover the binary characteristic of $\hat{\textbf{\text{x}}}^{(l)}$ as follows:
\begin{itemize}
	\item Reshape $\hat{\textbf{\text{x}}}^{(l)} \in [0, 1]^{(MN+N)}$ to its corresponding matrix form $\hat{\textbf{\text{X}}}^{(l)} \in [0, 1]^{N\times(M+1)}$;
	\item For each row of $\hat{\textbf{\text{X}}}^{(l)}$, set the element with the highest value to 1 and the others to 0 to obtain the matrix $\bar{\textbf{\text{X}}}^{(l)} \in \{0, 1\}^{N\times(M+1)}$.
\end{itemize}
Finally, the optimal solution to $\mathscr{P}$1 can be obtained by searching over all $L$ matrices $\bar{\textbf{\text{X}}}^{(l)}$.

The detailed algorithm is described in Algorithm 1 which can be considered as a variant of the randomization approaches in \cite{b3},\cite{b18},\cite{b19}. Note that the symbols $\text{z}^*_{ik}$, $\hat{\text{x}}^{(l)}_{ik}$ and $\bar{\text{x}}^{(l)}_{ik}$ represent the element in $i^\text{th}$ row and $k^\text{th}$ column of matrix $\textbf{\text{Z}}^*$, $\hat{\textbf{\text{X}}}^{(l)}$ and $\bar{\textbf{\text{X}}}^{(l)}$, respectively.

\begin{algorithm}[h]
\KwIn{M, N, L, $\alpha_i$, $\beta_i$, $\omega_i$, $r_k$, $C_k^{\text{UL}}$, $C_k^{\text{DL}}$;}
Solve $\mathscr{P}$1 without the rank constraint to obtain $\textbf{Z}^*$\;
Extract the $Q\times Q$ upper left sub-matrix $\textbf{\text{Z}}'$ from $\textbf{\text{Z}}^*$\;
\If{\text{rank}($\textbf{\text{Z}}^*$) = 1}
{
	the found objective function is minimal\;
}
\Else{
	$\gamma \leftarrow z^*_{Q+1, Q+3}$\;
	\For{l = 1 to L}{
		Generate $\hat{\textbf{\text{x}}}^{(l)} \sim \mathcal{N}(0_{(MN+N)\times1}, \textbf{Z}')$\;
		Reshape $\hat{\textbf{\text{x}}}^{(l)} \in \mathbb{R}^{MN+N}$ to $\hat{\textbf{\text{X}}}^{(l)} \in \mathbb{R}^{N\times(M+1)}$\;
		\For{i=1 to N}{
			$k^* \leftarrow \text{arg max}_{k \in \mathcal{M}}{\hat{\textbf{\text{x}}}}^{(l)}_{ik}$\;
			$\bar{\textbf{\text{x}}}^{(l)}_{ik^*} \leftarrow 1$ and $\bar{\textbf{\text{x}}}^{(l)}_{ik} \leftarrow 0, k \in \mathcal{M}\texttt{\symbol{92}}\{k^*\}$\;
	}
	Achieve $\bar{\textbf{\text{X}}}^{(l)}$\;
	}
	Find the smallest $\psi(\bar{\textbf{\text{X}}}^{(l)})$ over L matrices $\bar{\textbf{\text{X}}}^{(l)}$\;
}

\KwOut{Optimal cost function $\psi$ value}
\caption{{\bf SDR-based Task Offloading} \label{Algorithm}}
\end{algorithm}

\section{Simulation Results}
In this section, we investigate the performance of the proposed algorithm based on the parameters value specified in \cite{b2},\cite{b4},\cite{b20},\cite{b21}. Concretely, we consider a device with CPU frequency of 400$\times\text{10}^\text{6}$ Hertz and $P^\text{Comp}$, $P^{\text{Tx}}$, $P^{\text{Rx}}$ of 0.8W, 1.258W and 1.181W, respectively. We choose Gzip as the application to show the relationship between the number of computational cycles and input data size, that is, $\omega_i = \kappa \alpha_i$ where $\kappa =$ 330 cycles/bit, according to \cite{b20}. The output data size $\beta_i$ is set to be of 20\% of the input data size $\alpha_i$. The size of input data is $\alpha_i =$ 4Mb, $\forall i \in \mathcal{N}$. The uplink and downlink data rates are controlled in three ranges, from 500kbps to 1Mbps, 1Mbps to 2Mbps and 2Mbps to 10Mbps representing low, average and high data rate ranges, respectively. The weights are $\lambda_t = \lambda_e =$ 0.5. The parameters $J^{\text{C}}$ and $E^{\text{C}}$ are uniformly distributed within [200, 500] cycles/bit and [10, 20]$\times 10^{-11}$ Joules/cycle, respectively. The number of Gaussian samples $L$ = 100 and the number of CAPs is two, whose rates are 2$\times 10^9$ and 2.2$\times 10^9$ Hertz. The simulation results are obtained after averaging over 50 realizations.
\begin{figure*}[htbp]
	\center{\includegraphics[width=1\textwidth]{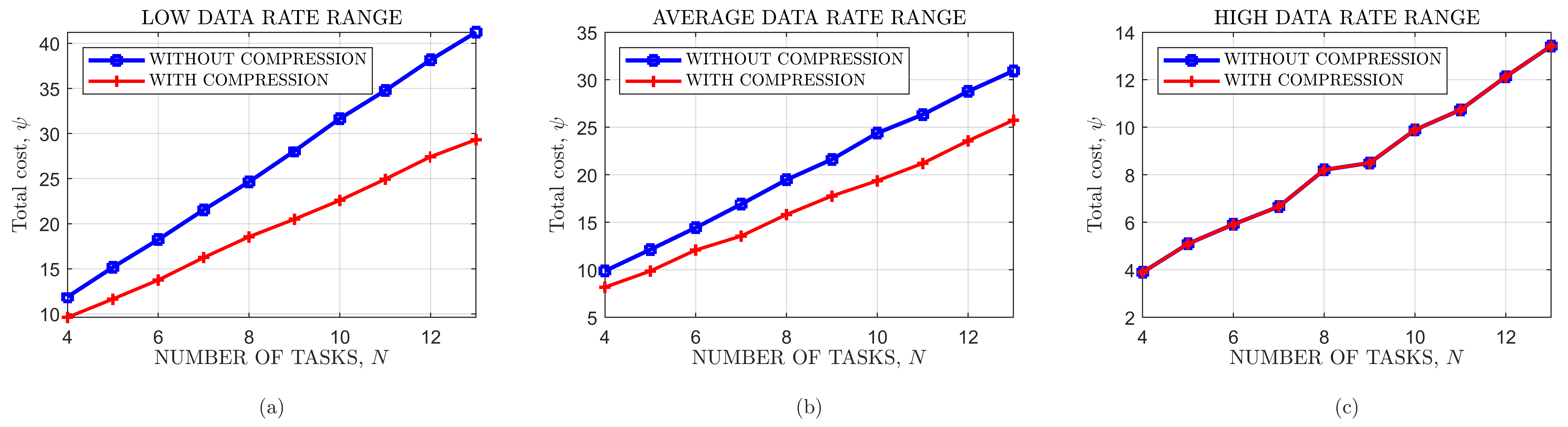}}
	\caption{Algorithm comparison when data rates are generated randomly (a) from 500kbps to 1Mbps (b) from 1Mbps to 2Mbps (c) from 2Mbps to 10Mbps.}
	\label{fig: Hinh2}
\end{figure*}
Fig. \ref{fig: Hinh2} compares the proposed scheme with the benchmark one where there is no data compression before the transmission. It is shown that the MD's energy and time response of the applications are saved the most when data rate is low. When the data rate increases, the total cost functions of the two algorithms are reduced as compared to the low data rate case, and the data compression step has less effect on the cost reduction. Finally, in case of high data rate range of [2, 10] Mbps, the two schemes output the same results when the MD decides that there is no need to compress the data before offloading because the amount of time response saved outweighs the amount of energy used for wireless transmission.

\section{Conclusion}
This work has investigated a computation offloading framework where one MD can offload tasks to different available CAPs. We have focused on minimizing the total cost which includes both the MD tasks' execution latency and energy consumption by coupling data compression and task allocation decisions. The non-convex problem was converted into a convex one through an SDR approach. Our simulations show that the data compression scheme is advantageous in cases of low data rates. In this work, we apply the same data compression ratio to all the offloaded tasks, which may not true in reality when the data sizes vary among different tasks. Thus, this scenario may serve as a fruitful avenue for future works.

\section*{Acknowledgement}
This research is funded by Vietnam National Foundation for Science and Technology Development (NAFOSTED) under grant number 102.04-2017.308.

\balance
\end{document}